# Spatial Dynamics of Invasion: The Geometry of Introduced Species

Gyorgy Korniss[a] and Thomas Caraco[b]

[a] *Department of Physics, Applied Physics and Astronomy, Rensselaer Polytechnic*

*Institute, Troy, New York 12180, U.S.A., email: kornig@rpi.edu*

[b] *Department of Biological Sciences, University at Albany, Albany, New York 12222,*

*U.S.A., email: caraco@albany.edu*

**Abstract**
Many exotic species combine low probability of establishment at each introduction with rapid population growth once introduction does succeed. To analyze this phenomenon, we note that invaders often cluster spatially when rare, and consequently an introduced exotic's population dynamics should depend on locally structured interactions. Ecological theory for spatially structured invasion relies on deterministic approximations, and determinism does not address the observed uncertainty of the exotic-introduction process. We take a new approach to the population dynamics of invasion and, by extension, to the general question of invasibility in any spatial ecology. We apply the physical theory for nucleation of spatial systems to a lattice-based model of competition between plant species, a resident and an invader, and the analysis reaches conclusions that differ qualitatively from the standard ecological theories. Nucleation theory distinguishes between dynamics of single-cluster and multi-cluster invasion. Low introduction rates and small system size produce single-cluster dynamics, where success or failure of introduction is inherently stochastic. Single-cluster invasion occurs only if the cluster reaches a critical size, typically preceded by a number of failed attempts. For this case, we identify the functional form of the probability distribution of time elapsing until invasion succeeds. Although multi-cluster invasion for sufficiently large systems exhibits spatial averaging and almost-deterministic dynamics of the global densities, an analytical approximation from nucleation theory, known as Avrami's law, describes our simulation results far better than standard ecological approximations.

*Keywords*: invasion criteria, nucleation theory, spatial competition

## 1. Introduction

The breakdown of biogeographic barriers allows some introduced species to reshape communities (Drake et al., 1989; Hengeveld, 1989; Rosenzweig, 2001) and threaten local biodiversity (Kolar and Lodge, 2002; Pimm, 1987), especially in nature reserves (Usher et al., 1988). Most introductions fail to initiate invasion (Lonsdale, 1999; Simberloff, 2000). However, an exotic's abundance often increases rapidly once introduction does succeed (Christian and Wilson, 1999; Sax and Brown, 2000; Shigesada and Kawasaki, 1997), particularly when an established exotic has an ecological advantage promoting its growth (Callaway and Aschehoug, 2000; Mack et al., 2000; Pimentel et al., 2000).

Veltman et al. (1996) analyze 496 documented, intentional introductions of 79 bird species to New Zealand. The strongest predictor of establishment is repeated introduction. Most species introduced four or fewer times never became established. Despite multiple introduction attempts for most species, only 20% of the birds ever became established (Veltman et al., 1996). Repeated failure of introduction, followed by ecological success once established, appears characteristic of both natural dispersal and human-mediated introduction (Sax and Brown, 2000). The seeming inconsistency between repeated failure of the introduction process and an invader's rapid growth once established motivates our study.

Despite the observed uncertainty of introduction, spatial models for invasion processes typically yield deterministic criteria for growth when rare (Andow et al., 1990; Caraco et al., 2002; Chesson, 2000; Kot et al., 1996; cf. Lewis and Pacala, 2000). An invader's growth or decline will ordinarily depend on locally structured interactions (Ellner et al., 1998; Wilson, 1998), so that chance mechanisms should often govern the population dynamics of rarity (Durrett and Levin, 1994a). Our results show how introduction rates and the size of the environment can generate random variation in an invader's success or failure, through effects on the invader's spatial clustering.

Our analysis specifically distinguishes single-cluster growth from multi-cluster growth of a rare exotic competing with a resident plant species through clonal propagation (Harada and Iwasa, 1994; Inghe, 1989). Simulating the model reveals interesting variation in the waiting time for successful introduction and subsequent spread of the exotic. To characterize our particular results and, more importantly, to offer a new perspective on the population dynamics of invasion, we invoke the physical theory for homogeneous nucleation of spatial systems (Avrami, 1940; Johnson and Mehl, 1939; Kolmogorov, 1937). Originally formulated to model processes such as crystallization, nucleation theory readily addresses ecological clustering generated by local propagation in viscous populations (Gandhi et al., 1999). We emphasize that under multi-cluster growth of the exotic species, the dynamics of competition for space, specifically the time-dependent decay of the resident's density, follows a powerful analytic approximation referred to as Avrami's law (Duiker and Beale, 1990; Ishibashi and Takagi, 1971; Ramos et al., 1999; Rikvold et al., 1994).

## 2. Spatial model for invader -resident competition

Individual plants typically interact more with nearby than with distant individuals (Rees et al., 1996; Tilman et al., 1997). Consequently, an introduction's success or failure can depend on effects regulated by neighborhood, rather than global, densities (Higgins et al., 1996; Wilson, 1998). We model two clonal species, a resident and an invader, competing for space in a lattice environment. Each of the $L^2$ lattice sites is either empty or occupied by a single plant; a site represents the minimal space an individual (ramet) requires. Competition for space is pre-emptive; a site already occupied cannot be colonized by either species until the occupant's mortality opens the site. Table 1 defines the model's symbols.

The elementary state of any site $x$ belongs to the set $\sigma = \{0, i, r\}$. The states indicate, respectively, an empty site, occupation by an individual invader, and occupation by an individual of the resident species.

First, we describe $(0 \to i)$ and $(0 \to r)$ transitions. An empty site may become occupied as a result of introduction from outside the environment, or through local clonal propagation (Cook, 1983; Iwasa, 2000). The invader occupies each empty site via dispersal at constant probabilistic introduction rate $\beta_i$. The introduction rate for the resident species is $\beta_r$. The introduction process does not depend on an open site's local neighborhood; introduction corresponds to a spatially uniform, typically weak, background process, modeling long-distance propagule dispersal.

Local clonal propagation depends on neighborhood composition. A plant occupying a site $x$ may propagate locally if at least one of the $\delta$ sites neighboring $x$ is open. An invader at site $x$ attempts to colonize neighboring sites at total probabilistic rate $\alpha_i$; the propagation rate per neighboring site is $\alpha_i/\delta$. The resident's total propagation rate is $\alpha_r$, so the rate per site is $\alpha_r/\delta$. The chance of successful clonal growth declines with local density. $\eta_i(x, t)$ counts invaders neighboring an open site $x$ at time $t$. $\eta_r(x, t)$ counts individuals of the resident species on the same neighborhood; $\eta_i(x, t) + \eta_r(x, t) \le \delta$.

We assume density-independent mortality. If a site is occupied by an invader, that site becomes open at constant probabilistic rate $\mu_i$. The mortality rate for individuals of the resident species is $\mu_r$.

Next we specify the model's transition rates. Introduction and local propagation occur independently, so that an open site becomes occupied by an invader, a $(0 \to i)$ transition, at total rate $\beta_i + \eta_i(x, t) \, \alpha_i/\delta$. An open site becomes occupied by a resident, a $(0 \to r)$ transition, at total rate $\beta_r + \eta_r(x, t) \, \alpha_r/\delta$. The $(i \to 0)$ transition, where invader mortality opens a site, occurs at rate $\mu_i$; the $(r \to 0)$ transition occurs at rate $\mu_r$.

Throughout, we set $\beta_i = \beta_r = \beta$, and $\mu_i = \mu_r = \mu$, so that asymmetry between the invader and resident is due solely to the difference in the local propagation rates $\alpha_i$ and $\alpha_r$. We keep $\alpha_i > \alpha_r$, and ask how the invader's clonal-growth advantage affects the dynamics of invasion as we vary lattice size and the introduction rate. We keep $\beta$ much smaller than the other rates, so that neighborhood competition drives the dynamics, but trapping states are avoided during simulation.

## 3. Mean-field approximation

The mean-field approximation offers a clear picture of the basic biological problem, and so guides analysis of the spatial model. The mean-field model assumes homogeneous mixing, and so ignores effects of spatial clustering on the dynamics.

Let $\rho_j$ represent the global density of sites with state $j$. Under the mean-field approximation,

$$d\rho_j/dt = \left(\beta + \alpha_j \rho_j\right)\left(1 - \rho_r - \rho_i\right) - \mu \rho_j; \quad j = r, i \qquad (1)$$

To focus on spatial competition, we temporarily suppress immigration by setting $\beta = 0$ (see Lehman and Tilman, 1997). Then there are three equilibrium fixed points $(\rho^*_r, \rho^*_i)$; none allows coexistence. Mutual extinction, $(\rho^*_r, \rho^*_i) = (0, 0)$, is stable when each $\alpha_j < \mu$. Competitive exclusion of the invader by the resident, where $(\rho^*_r, \rho^*_i) = (1 - \mu/\alpha_r, 0)$, is stable when $\alpha_r > \mu$ and $\alpha_r > \alpha_i$. Finally, exclusion of the resident species by the exotic, where $(\rho^*_r, \rho^*_i) = (0, 1 - \mu/\alpha_i)$, is stable when $\alpha_i > \mu$, and $\alpha_i > \alpha_r$. We assume $\alpha_i > \alpha_r > \mu$, so that each species can grow in an empty environment, and the invader has an individual-level advantage. Given this ordering of parameters, only the last stability criterion can hold. Therefore, under homogeneous mixing, the exotic is introduced at rate $\beta$ and the invasion condition, i.e. the condition for the exotic's advance when rare against the resident at positive equilibrium, is simply $\alpha_i > \alpha_r$. Since the invader has this advantage, it will proceed to exclude the resident (to order $\beta$ with introduction allowed) under homogeneous mixing.

Integrating Eqs. (1) numerically yields time dependent global densities. Figure 1 shows the results for different values of $\beta$ ($\beta << \mu, \alpha_r, \alpha_i$). Initializing the environment as fully occupied by resident, the system very rapidly relaxes to a phase dominated by the resident species, where the global densities are simply $(\rho^*_r, \rho^*_i) = (1 - \mu/\alpha_r, 0)$ up to order $\beta$. This fixed point of Eqs. (1) is not stable, but the densities stay very close to these values for some time before the invader's growth becomes noticeable. In this sense, we refer to this phase where the resident resists invasion as "meta-stable." We define the characteristic time $\tau$, the lifetime of the resident species, as the time until the resident's density decays to half of its meta-stable value. As $\beta$ is decreased, the global densities as functions of time shift to the right (Fig. 1); note that the increase of the lifetime as $\beta$ decreases is extremely slow, proportional to $-\log(\beta)$.

As noted above, approximating the spatial dynamics by mean-field or pair correlation methods (Bolker and Pacala, 1999; Matsuda et al., 1992), yields deterministic criteria for advance when rare; these methods, by construction, cannot address the observed stochastic variation in success or failure of introduction. Furthermore, for parameter values we use in simulations of the spatially detailed model, the standard approximations fail to describe the dynamics following invasion; see Discussion. After describing our simulations, we apply nucleation theory to characterize both the uncertainty of the introduction process and the advance of the invader.

## 4. Simulation results

We implemented standard dynamic Monte Carlo (MC) simulations on an $L \times L$ lattice with periodic boundaries. Neighborhood size was $\delta = 4$. The time unit was one MC step per site (MCSS), during which $L \times L$ sites were picked randomly and updated probabilistically. This procedure simulates continuous-time dynamics in the large $L$ limit (Durrett and Levin, 1994b; Korniss et al., 1999). From the mean-field computations, one might not expect the competitively inferior resident to resist invasion and induce slow dynamics, since the lifetime of the resident species increased only logarithmically with decreasing $\beta$. However, simulation of the spatial model revealed slow "meta-stable" decay of the resident. That is, spatially structured interactions slow the timescale of competitive systems (Hurtt and Pacala, 1995; Lehman and Tilman, 1997).

Fixing $\mu = 0.10$, $\alpha_r = 0.70$, and $\alpha_i = 0.80$ (the same values used in the mean-field approximation), we found qualitative variation in the dynamics for introduction rates $10^{-8} \le \beta \le 10^{-4}$, and similarly for system sizes $32 \le L \le 512$. At $t = 0$ we initialize the environment with all sites occupied by the resident. After about (typically less than) 10 MCSS the system relaxes to a "meta-stable" configuration dominated by the resident with a small density of empty sites; we designate the meta-stable density of the resident as $\rho_r = r_{ms}$. Individuals of the exotic species occasionally occupy empty sites via introduction. An immigrant invader may die without propagating. If a site opens up in the neighborhood of a rare invader, the empty site is likely surrounded by more than one resident. The resident's greater local density more than compensates for its lower propagation rate per individual, so the resident has the better chance of colonizing the empty site; see Discussion. Consequently, small clusters of invaders usually shrink and disappear. Introductions fail since preemptive competition imposes a strong constraint on the exotic's growth. Introduction succeeds only if the exotic can generate a spatial cluster large enough that it statistically tends to grow at its periphery. On coarse-grained length scales, nucleation theory suggests that there exists a critical radius for invader clusters; smaller clusters decline in size, while clusters with a radius larger than the critical value will more likely grow than decline (Gandhi et al., 1999). Simulations confirm this picture, and suggest an interesting distinction between single and multi-cluster invasion. For $\beta = 10^{-6}$, a $128 \times 128$ system exhibits single-cluster invasion (Fig 2). Increasing $\beta$ to $10^{-4}$ produces multi-cluster invasion (Fig. 3). Holding $\beta$ constant and increasing the system size $L$ would also generate multi-cluster invasion. Since the spatial structure of the introduction process exhibits nucleation and growth of invader clusters, we apply nucleation theory as described in the next section.

## 5. Nucleation theory

Nucleation theory's ecological significance lies in its prediction of population dynamics according to an invader's spatial-clustering pattern. Our application of nucleation theory rests on the conceptual similarity between biological invasion and meta-stable decay in physical systems. In particular, our analysis parallels the theory developed by Rikvold et al. (1994), Richards et al. (1995), and Korniss et al. (1999) for magnetization switching in ferromagnetic materials.

When the resident species' density $\rho_r(t)$ first falls to $r_{ms}/2$, we consider the resident competitively dominated by the exotic (defining competitive dominance in terms of any particular global fraction of sites which the resident occupies does not affect the dynamics). The non-negative random variable $\tau$ will represent the first-passage time of the resident's density to $r_{ms}/2$, in accordance with earlier definition of the lifetime for the mean-field approximation. The mean waiting time $\langle \tau \rangle$ is called the meta-stable lifetime; the expected time elapsing until the resident's global density is ½ its meta-stable value. In single-cluster invasion $\tau$ sums the random waiting time until successful introduction and a subsequent period of invasive spread. Nucleation is equivalent to successful introduction; invasive spread begins when a critically sized cluster of invaders first forms. In multi-cluster invasion, clusters of the critical size continuously form and grow, leading to many invading species' clusters of various sizes; see snapshots of the environment in Fig. 3. In this section we identify how the probability distribution of $\tau$ differs between single-cluster and multi-cluster invasion, and interpret this difference ecologically.

### 5.1. Single-cluster invasion

In single-cluster invasion, random variation in the time until a cluster exceeds critical size contributes more to the variance of $\tau$ than does variation in the period of invasive growth following successful introduction. Different realizations of the introduction and invasion dynamics (simulations with the same initial configuration of the resident, but using different random numbers) show that the average time for meta-stable decay has little predictive significance for single-cluster dynamics. In fact, the standard deviation is comparable to the average (Fig. 4a); the ecological implication is that uncertainty of the success or failure of introduction makes single-cluster invasion inherently stochastic.

When the typical cluster separation (of a hypothetical infinite system) exceeds the finite environment size $L$, the spatial dynamics of the latter (finite) system almost always exhibits meta-stable escape through nucleation and growth of a single cluster of the superior species (Rikvold et al., 1994). Ecologically, if the introduction rate or environment size is sufficiently small, the invasion geometry will always involve single-cluster dynamics. While the introduction of an invader at an empty site is a Poisson process by the construction of our stochastic spatial model, it is not known *a priori* whether the nucleation of a successful invading cluster (one which has just reached the critical size) will also be Poisson. However, we shall suppose, as a working assumption, that nucleation of a single invader cluster is a Poisson process as well, and verify this assumption using simulations. Following this assumption, single-cluster invasion (advance of the first critical cluster) should be described by an exponential distribution of time until successful introduction of the invader. Consider the cumulative probability distribution of competitive waiting times, i.e., the probability that the resident's global density has not decayed to $r_{ms}/2$ by time $t$, $P_{not}(t) = P(\tau > t)$. For single-cluster invasion, analysis by Richards et al. (1995) shows us that

$$P_{not}(t) = \begin{cases} 1 & \text{for} \quad t < t_g \\ \exp\left(-\dfrac{t - t_g}{\langle t_i \rangle}\right) & \text{for} \quad t > t_g \end{cases} \tag{2}$$

where $\langle t_i \rangle$ is the mean time until successful introduction of the exotic (mean nucleation time) and $t_g$ is the duration of invasive spread; we take the latter as constant for single-cluster invasion, since the growth of a single, supercritical cluster can be treated deterministically. Note that $t_g$ does not depend on $\beta$, but only on $\mu$, $\alpha_i$, $\alpha_r$, and system size $L$. Given the symmetry of resident decline and invader growth (Fig. 4a), we approximate $t_g$ as the time for the supercritical cluster of the invading species to grow and fill half the environment. For fixed $\mu$, $\alpha_i$, and $\alpha_r$, the characteristic time scale (the mean nucleation time $\langle t_i \rangle$), depends only on $\beta$ and the system size $L$. More precisely, nucleation of a critical cluster is a Poisson process with nucleation rate per unit area $I(\beta)$. Thus, for systems in the single-cluster regime

$$\langle t_i \rangle \propto [L^2 I(\beta)]^{-1} \tag{3}$$

in two dimensions (Rikvold et al., 1994). The distribution of first-passage time $\tau$ has a two-parameter exponential density (see Bury, 1975) with mean $\langle \tau \rangle = (\langle t_i \rangle + t_g)$, variance $\langle t_i \rangle^2$, and skew $2\langle t_i \rangle^3 > 0$.

To verify that nucleation of a cluster of critical size is a Poisson process, we simulated $10^3$ ($10^4$ for some parameter values) independent realizations of introduction and invasion, and constructed the cumulative distribution $P_{\text{not}}(t)$. Figure 5, (a) and (b), compares the results with Eq. (2) for fixed $\beta$ and various $L$ values, and for fixed $L$ and various $\beta$ values, respectively. Fitting an exponential to the data (a straight line on log-linear scales), we estimated the mean time to successful introduction $\langle t_i \rangle$ (i.e., the nucleation time of the first critical cluster) as the inverse of the slope in Fig. 5. Then one can read the invasive-spread time $t_g$ from the figure; from expression (2) $t_g$ is equal to the maximal time at which $P_{\text{not}}(t)$ is unity.

We also confirmed the dependence of $\langle t_i \rangle$ on $L$, as indicated in Eq. (3), and found that $I(\beta) \propto \beta$; the rate of successful introduction is proportional to the introduction rate. Since the typical separation between invading species' clusters increases with decreasing $\beta$ (Richards et al., 1995), the fundamental consequence of these findings is that given an arbitrarily large but finite and fixed system size $L$, for sufficiently small $\beta$ the system will exhibit single-cluster growth of the invader with $\langle t_i \rangle \propto \beta^{-1}$. Note that this behavior of the time elapsing until successful introduction follows directly follows from the microscopic dynamics for small $\beta$. However, $\langle t_i \rangle$ remains orders of magnitude larger than one would estimate based simply on the system size and the density of open sites, since many exotic clusters fail to grow before introduction succeeds. Most importantly, the average lifetime for decay of the resident's density increases rather quickly as $\beta$ decreases, in stark contrast with the weak logarithmic divergence predicted by the mean-field model (and pair approximation models; see Discussion). Also note that for small $\beta$, $\langle \tau \rangle \approx \langle t_i \rangle$, (since $\langle t_i \rangle >> t_g$), and $\langle t_i \rangle$ equals to the standard deviation of the exponential lifetime distribution, Eq. (2).

When introduction rates or the numbers of habitable sites in an environment are sufficiently small, ecological invasion of a locally propagating species should occur as the growth of a single invader cluster. The inherent stochasticity of the single-cluster process renders invasion time highly unpredictable. The standard deviation of the meta-stable lifetime $\tau$ is as large as the mean, and standard ecological methods for spatial systems do not address the uncertainty of invasion dynamics. Nucleation theory, as a first step, provides a functional form characterizing the random waiting time of single-cluster invasion process.

### 5.2. Multi-cluster invasion

In multi-cluster invasion, realizations of introduction and invasion processes are quite different from those of single-cluster growth. Species' global densities exhibit only small fluctuations about their time-dependent averages (Fig. 4b; cf. Fig. 4a). Therefore, the first-passage time of the resident's density to half its initial value has a standard deviation much smaller than its average $\langle \tau \rangle$. Nucleation and subsequent expansion of many clusters implies that global density is a sum of random variables, and spatial averaging of local densities within the multi-cluster process reduces the variability of the time-dependent global densities among different realizations of the process. This reduction in variability of the dynamics of the global species' densities, resulting from averaging local behavior of a large number of clusters, is sometimes termed "self-averaging."

To model decay of the resident's density under multi-cluster invasion, we again invoke nucleation theory. Increasing the introduction rate $\beta$ or system size $L^2$ leads to the nucleation and subsequent growth of many invader clusters (Ramos et al., 1999). The meta-stable lifetime $\langle \tau \rangle$, the resident's mean first-passage time to $r_{ms}/2$, becomes independent of the size of the environment; the standard deviation is proportional to the inverse of the square root of system size, i.e., $1/L$. Further, since global densities reflect spatial averaging of local densities during the nucleation and growth of many invader clusters, the distribution of the first-passage time $\tau$ is normal (Richards et al., 1995). We express the corresponding cumulative probability distribution $P_{not}(t)$ as an error function (Fig. 6a); the point where the $P_{not}(t)$ functions for different system sizes cross corresponds to the system-size independent meta-stable lifetime for the resident species' decay (Korniss et al., 1999). Compared to the single-cluster mode, the mean first-passage time $\langle \tau \rangle$ is decreased (and becomes system-size independent in the large-$L$ limit). Importantly, as the size of the environment invaded increases, the variance and skew of $\tau$ go to zero.

The preceding observations imply that, for multi-cluster invasion, global densities $\rho_j(t)$ converge to deterministic functions in large environments. Hence we tested KJMA theory, so named for its developers: Kolmogorov (1937), Johnson & Mehl (1939) and Avrami (1940). In particular, we tested Avrami's law, which predicts the time-dependent global density of the resident species when the superior competitor invades through the nucleation and growth of many clusters. The Avrami picture of meta-stable decay works accurately until invader clusters begin to coalesce, and the invader becomes the more abundant species; Ramos et al. (1999) offer a visual characterization of multi-cluster dynamics. According to Avrami's law the global density of the meta-stable resident species decays as

$$\rho_r(t) \approx r_{ms} \exp\left[-\ln 2\left(\frac{t}{\langle \tau \rangle}\right)^3\right] \tag{4}$$

where $\langle \tau \rangle$ is the asymptotically system-size independent mean lifetime of the resident's decay. The Appendix presents a derivation of Avrami's law based on the assumptions of multi-cluster nucleation and applies the general result to our ecological-invasion problem to obtain expression (4). Figure 6(b) shows convincing agreement between simulation averages and Avrami's law; deviations are noticeable only for very large times [$t^3 \geq 10^{10}$ ($t \geq 2154$), inset of Fig. 6b] when invader clusters begin to coalesce (Korniss et al., 1999) and percolation effects become important (Gandhi et al., 1999).

Nucleation theory also relates the system-size independent lifetime $\langle \tau \rangle$ of the multi-cluster regime to the inherent nucleation rate per unit area $I(\beta)$ through

$$\langle \tau \rangle \propto \left[I(\beta)\right]^{-1/3} \tag{5}$$

[see Eq. (A.6) of the Appendix]; note that this proportionality is quite different than that obtained above for single-cluster dynamics. Eq. (5) enables us to predict the $\beta$-dependence of the lifetime in the multi-cluster regime based on measuring the mean nucleation time in single-cluster invasions; see Eq. (3). Combining that result with Eq. (5) implies that $\langle \tau \rangle \propto \beta^{-1/3}$ in the multi-cluster regime. Thus, for an arbitrary small $\beta$, in the limit as $L \to \infty$, the lifetime of the resident species' decline eventually approaches an asymptotic system-size independent value, proportional to $\beta^{-1/3}$. While the global densities become deterministic in the multi cluster regime, the $\beta$-dependence of the lifetime is very different from the weak logarithmic divergence yielded by the mean-field approximation (pair approximation, not shown, is similar to the mean field). That is, invader density may vary locally across the environment, but the multi-cluster dynamics will maintain the spatially averaged, i.e. the global, density with little variability about a time-dependent mean. Nevertheless, despite this lack of random variation in the global densities, the multi-cluster dynamics with nearest-neighbor interaction may be poorly predicted by standard approximations to spatial processes. But Avrami's law, Eq. (4), accurately predicts the simulated multi-cluster invasion dynamics.

## 6. Discussion

Theory in population biology relies heavily on invasion analyses. That is, predictions commonly are inferred from conditions promoting the initial increase of a rare type (allele, phenotype, or species) among resident types, usually at dynamic equilibrium (see Ferriere and Gatto, 1995). Invasion criteria formalize conditions for local stability of the rare type's extinction; when extinction is unstable, invasion succeeds. Models for spatial population processes may be deterministic or stochastic, but the associated invasion criteria typically are deduced from linear, deterministic approximations to a rare type's dynamics. However, most introductions in nature presumably begin with a small number of colonists, implying that a more realistic approach to invasion analysis would model discrete individuals and include a random component in the dynamics of rarity (Caraco et al., 1998; Durrett and Levin, 1994a). Nucleation theory characterizes stochastic properties of introduction and invasion that must often result from spatially structured interactions and locally clustered growth.

Our analysis associates random variation in the spatial dynamics of an introduced species with the contrast between the high probability that a given introduction will fail and the ecological dominance of exotics once introduction succeeds (Kolar and Lodge, 2002; Sax and Brown, 2000). If the exotic's introduction rate or the system size is small, invasion occurs almost always through a single successful invading cluster. Before that successful event, invader clusters fail to achieve critical size and decline to extinction. The ultimate decline of the resident occurs only after a number of stochastic introductions of the exotic species fail; when the exotic invades successfully, it grows as a single super-critical cluster (Rikvold et al., 1994).

Multi-cluster invasion has distinct features. In larger systems, or with higher introduction rates, invasive growth of the exotic begins (almost) as soon as biogeographic barriers break down, and dynamics of the global densities becomes nearly deterministic. For multi-cluster invasion, nucleation theory approximates the decay of the resident by Avrami's law (Ramos et al., 1999; Richards et al., 1995). The global densities become self-averaging with an asymptotically system-size independent mean lifetime $\langle \tau \rangle$ for the resident. Quantitative or qualitative variation in local interactions, through effects on cluster formation and dissolution (Gandhi et al., 1999; van Baalen and Rand, 1998), influences $\langle \tau \rangle$ and consequently should exert predictable effects on global dynamics; see the Appendix.

Ecological implications of nucleation theory extend beyond the neighborhood-level competition we model. The two limiting processes we emphasize suggest general significance for spatial ecologies. Consider a hypothetically infinite system (i.e., when we take the limit $L \to \infty$ first). Then, for all $\beta$ much smaller than $\alpha_i$, $\alpha_r$, and $\mu$ (so that competition drives the dynamics), the system must be in the multi-cluster regime. Spatial averaging of local dynamics generates (almost) deterministic behavior of the global densities, and Avrami's law accurately describes the resident's decay. Although the global densities are deterministic functions of time for large systems, recall that their dynamics qualitatively differ from the results of the mean-field approximation (Figs. 7a and b). Further, the lifetime $\langle \tau \rangle$ increases as $\beta^{-1/3}$ for decreasing $\beta$, much faster than the weak logarithmic increase predicted by the mean-field model (Fig. 7c).

We also developed a pair approximation to the spatially detailed model (after Iwasa et al., 1998). Pair approximation incorporates short-range spatial correlations, and so evaluates both global densities and the conditional densities of the states of paired, neighboring sites (Rand, 1999). The pair approximation marginally improves the estimation of the resident species' lifetime over the mean-field model; pair approximation might better predict global densities for interaction neighborhoods extending beyond nearest neighbors (e.g., Caraco et al., 2001). Importantly, both mean-field and pair approximations fail to capture the actual behavior of the time-dependent global densities of the spatial model, well described by nucleation theory and Avrami's law.

Now consider the second limiting process. For any finite system, there is a sufficiently small $\beta$ where the typical cluster separation increases beyond the system size, and invasion crosses over to the single-cluster mode. Below this $\beta$, we found that $\langle \tau \rangle \propto \beta^{-1}$; this is the $\beta \to 0$ limiting behavior of the meta-stable lifetime for a finite system. Illustrating this scenario for an $L = 128$ system (Fig 7c), we confirmed that for $10^{-5} \le \beta \le 10^{-3}$ the system is in the multi-cluster regime and the $\beta$-dependence of the lifetime follows $\beta^{-1/3}$. At around $\beta \approx 10^{-6}$ the crossover occurs, and for $\beta \le 10^{-6}$ the system exhibits single-cluster invasion and indeed approaches the $\langle \tau \rangle \propto \beta^{-1}$ behavior (Fig. 7c). Here the invasion is inherently stochastic, and the mean of the lifetime becomes equal to its standard deviation. Note that in the large-$\beta$ region ( $\beta \ge 10^{-2}$ ) nucleation theory breaks down as invading clusters coalesce almost immediately after introduction. In fact, the mean-field and pair approximations begin to work much better here as a result of the almost immediate mixing of small clusters.

We set $\beta_i = \beta_r$ for simplicity; the difference between species was due solely to different rates of local propagation. However, equal introduction rates might imply that the competitors' populations outside the environment are the same size (or equally distant), which is unlikely. For the initial conditions we consider, however, the dynamics is insensitive to $\beta_r$ (as long as it remains small), since the system is initially occupied densely by the resident species. Assigning separate introduction rates to the species, in particular, $\beta_i \ll \mu, \alpha_r, \alpha_i$ as previously, but with $\beta_r = 0$, does not alter the dynamics qualitatively; the ecological distinction between single-cluster and multi-cluster invasion processes remains important.

### 6.1. Local dynamics

Our results, in particular the slow decay of the resident species' density, depend on the discreteness of space. Bolker et al. (2000) suggest that lattice-based models overemphasize effects of local clumping on dynamics, but we can regard a lattice site as the minimal amount of space (hence, minimal access to resources) necessary to sustain an individual (ramet). At some scales, this may imply greater realism for the discrete-space approach (Neuhauser and Pacala, 1999). Of course, our detailed simulation results further depend on choices of parameter values, neighborhood size, and system sizes (Filipe and Maule, 2003; McCauley et al., 1993).

The local dynamics help explain the resistance to invasion. In the absence of the invader, the resident's global density will be approximated by $\rho^*_r = (1 - \mu/\alpha_r)$. Suppose that an individual of the invasive species is introduced, via dispersal, at an open site $y$ at time $t$. Approximating the local state frequencies by global densities (e.g., Duryea et al., 1999), the probability that no site is open (for local propagation) among the neighbors of site $y$ is close to $(\rho^*_r)^\delta$. If no site neighboring the single invader is open, the chance that the immigrant invader dies before a neighboring site opens is just $(\delta + 1)^{-1}$, since each individual has the same exponentially distributed waiting time for mortality. Larger neighborhoods increase the probability that the immigrant will find a neighboring site open, but do not necessarily increase the chance that the invader will propagate into an open site. To examine the latter probability, we consider a simple example.

Suppose that neighboring sites $(x, y)$ form an (open, invader) pair. The open site $(x)$ becomes occupied by the invading species at constant probabilistic rate $(\beta + \alpha_i / \delta)$. The same open site becomes occupied by the resident at rate $(\beta + \eta_r(x, t) \ \alpha_r / \delta)$, which depends on the number of resident individuals neighboring site $x$. Approximating that number $\eta_r(x, t)$ with a binomial random variable, the resident species occupies the open site at constant probabilistic rate $(\beta + (\alpha_r - \mu) \ (\delta - 1)/\delta)$. Although $\alpha_i > \alpha_r$, that resident is more likely to acquire the open site as long as

$$(\alpha_r - \mu)(\delta - 1) > \alpha_i \qquad (6)$$

This expression holds in our simulations; more generally, the invader is less likely to acquire the open site as neighborhood size $\delta$ increases.

The preceding emphasizes that it is the discreteness of the introduction process and the preemptive nature of the competition that, in combination, allow the resident species to repel small, rare clusters of invaders, before the resident declines. Recall that this is the sense in which we refer to the possibly lengthy domination of the environment by the resident species as "meta-stable." In contrast to Gandhi et al. (1999; see below), we do not assume an underlying bistability (such that either species, when common, tends to repel invasion by the other). That is, our initial global densities are unstable, and the system eventually reaches the stable stationary state dominated by the competitively superior invader.

### 6.2. Nucleation theory in biology

Gandhi et al. (1998, 1999) model two species competing for space, and consider how cluster growth/decay can influence time to extinction. The authors assume a somewhat elaborate locally structured dynamics. The mortality rate for individuals of each species increases as the total density in the local neighborhood increases. The birth rate for individuals of either species increases as the relative frequency of conspecifics increase locally. Finally, individuals may move diffusively (Gandhi et al., 1998).

As an initial condition, Gandhi et al. (1998, 1999) distribute large numbers of individuals of each species uniformly across the environment. Gandhi et al. (1998) assume competitively identical species. If initial densities differ sufficiently, a mean-field model approximates the time elapsing until the less numerous species reaches extinction. But when both competitors initially occur at high density, clusters quickly form in each species. Thereafter, the dynamics are driven by interactions at cluster interfaces, and the expected time to extinction increases beyond the mean-field prediction. Gandhi et al. (1999) conduct a similar analysis with asymmetric species, and they invoke nucleation theory's concept of a cluster size critical for growth in a competitive environment. The model's positive frequency-dependent birth rates likely accentuate the rapid decay of small clusters; smaller clusters have greater perimeter curvature, so that individuals on the perimeter have relatively few conspecific neighbors. Again, in their models nucleation is driven by an underlying bistability (two stable fixed points in terms of the global densities).

Our analysis complements the results in Gandhi et al. (1998, 1999). Each model addresses competition; we assume clonal propagation, while Gandhi et al. model social interactions where paired conspecifics do better than heterospecific pairs (see Giraldeau and Caraco, 2000). Our analysis focuses on ecological invasion, a process where one species begins at zero density, enters the environment as rare individuals, and advances through single or multi-cluster growth (depending on parameter values). Gandhi et al. (1998, 1999) initially disperse each species uniformly. Both species form multiple clusters, and then the sometimes lengthy process of competitive exclusion begins. Our model's biological assumptions differ considerably from the model by Gandhi et al. (1998, 1999), but both analyses point to the importance of cluster geometry as a basis for understanding relationships between individual-based interactions and global dynamics.

The term "nucleation" has been applied, metaphorically, in studies of community succession (Franks, 2003; Moody and Mack, 1988) and ecological restoration (Robinson and Handel, 2000). These applications refer to an interspecific facilitation, where a plant of one species modifies local sites in a manner promoting the germination and survival of a second species. Yarranton and Morrison (1974) describe an interesting example by tracing primary succession in a sand-dune community. Persistent vegetation (oak-pine forest) replaces colonizing grassland during succession. Persistent species begin as local clusters that expand and eventually coalesce, replacing the colonizing species in the process. But many of the persistent-species clusters are initiated through seedling establishment under individual junipers (*Juniperus virginiana*), where microclimate is moderated and soil nutrients are concentrated (Yarranton and Morrison, 1974). Our model assumes equivalent sites, but logical extensions of the theory would examine effects of spatial heterogeneity in site quality and exogenous temporal variation in demographic parameters (Korniss et al., 2001).

Finally, we suggest that nucleation theory offers a quantitative context for addressing a diverse series of fundamental questions in biology. For example, Herrick et al. (2002) analyze DNA replication by invoking formal equivalence between KJMA theory and their stochastic model for replication kinetics of *Xenopus* DNA. Initiation of replication forks along the linear DNA molecule is equivalent to nucleation events, and replication fork velocity is equivalent to the rate of cluster growth. Herrick et al. (2002) present the first reliable description of temporal organization in a higher eukaryote's DNA replication; nucleation theory provides the study's conceptual perspective and so guides the interpretation of data.

## Acknowledgements

Discussion with M. A. Novotny and Z. Toroczkai is gratefully acknowledged. We also thank Z. Toroczkai for helping us implement a numerical integrator routine for the mean-field and pair approximations. We appreciate the support of NSF Grant DEB-0342689. G. K. is also supported by NSF through Grant DMR-0113049 and by the Research Corporation through Grant No. RI0761.

**Appendix. KJMA theory for homogeneous nucleation**

Here we present a brief derivation of KJMA theory (commonly referred to as Avrami's law) for an infinite system. Avrami's law was originally formulated to describe how solids transform from one state of matter ("phase") to another as they crystallize. Since then, nucleation theory and Avrami's law have also been applied successfully to domain switching in ferromagnetic (Ramos et al., 1999; Rikvold et al., 1994) and ferroelectric (Duiker and Beale, 1990; Ishibashi and Takagi, 1971) materials. Our development follows Ishibashi and Takagi (1971). See Duiker and Beale (1990) for discussion of large, but finite systems, in particular, for consideration of finite-size effects where clusters begin to coalesce. The appendix first offers a description applicable to spatial systems in general, and then specifies application to the model analyzed in the text.

The system is initialized in the meta-stable phase. In homogeneous nucleation the decay of the meta-stable phase (or "switching" to the final equilibrium phase) occurs through random nucleation and subsequent growth of local clusters. Consider an arbitrary point $Q$ in the $d$-dimensional, infinite "volume." The probability that this point is not in the switched volume by time t, $P_{not}(t)$, equals to the volume fraction of the initial phase, $\varphi(t)$.

Recall that we expect an invading cluster to continue to grow only after its radius reaches a critical length $r_c$. We assume that nucleation of a successful invading cluster (with initial radius $r_c$) is a Poisson process with constant nucleation rate $I$ per unit volume (i.e., the probability of nucleation per unit volume per unit time is $I$). Such a cluster, nucleated at time $t'$, will cover a volume

$$S(t,t') = C_d [r_c + v(t-t')]^d \tag{A.1}$$

at later time $t$, where $v$, the radial velocity of a growing cluster, is approximated by a constant, and $C_d$ defines the relationship between radius and volume in the $d$-dimensional volume (i.e., $C_2 = \pi$ and $C_3 = 4\pi/3$).

Now divide the $(0,t)$ time interval into $N$ infinitesimal intervals $(j\Delta t, (j+1)\Delta t)$ with $\Delta t = t/N$, $j = 0, 1, 2, \ldots, N-1$. For infinitesimal $\Delta t$, under the assumption of a Poisson process, the probability that no cluster, nucleated in the infinitesimal interval $(j\Delta t, (j+1)\Delta t)$, will cover point $Q$ at time $t$ is $1 - IS(t, j\Delta t)\Delta t$. Thus, the probability that point $Q$ is not swept by *any* cluster (i.e., $Q$ is not in the switched volume) at time $t$ is given by

$$P_{not}(t) = [1 - IS(t,0)\Delta t][1 - IS(t,\Delta t)\Delta t][1 - IS(t,2\Delta t)\Delta t]\ldots[1 - IS(t,N\Delta t]$$
$$= \prod_{j=0}^{N-1}[1 - IS(t, j\Delta t)\Delta t]. \tag{A.2}$$

Taking the logarithm of Eq. (A.2), letting $\Delta t \to 0$, and letting $N \to \infty$ (such that $N\Delta t = t$) yields

$$\ln P_{not}(t) = \sum_{j=0}^{N-1} \ln[1 - IS(t, j\Delta t)\Delta t] \xrightarrow{\Delta t \to 0} -I\int_0^t S(t,t')dt' = -IC_d \int_0^t [r_c + v(t-t')]^d \, dt'$$
$$= \frac{IC_d}{v(d+1)}[r_c + v(t-t')]^{d+1}\Big|_0^t = -\frac{IC_d}{v(d+1)}\Big[(r_c + vt)^{d+1} - r_c^{d+1}\Big]. \tag{A.3}$$

In our case, as in many applications, the critical radius is much smaller than the typical cluster separation (which equals to the average diameter of the cluster when they begin to coalesce) and can be neglected. Then, for $r_c = 0$, the volume fraction of the meta-stable phase becomes

$$\varphi(t) = P_{not}(t) = \exp\left(-\frac{IC_d v^d}{d+1} t^{d+1}\right), \tag{A.4}$$

which is the general form of Avrami's law.

For our application, $d = 2$. The meta-stable phase corresponds to the competitively inferior, resident species with a small density of open sites in the background. The introduced species advances through nucleation (successful introduction) and subsequent cluster growth. Thus, for the decay of the resident we find

$$\rho_r(t) = r_{ms}\varphi(t) = r_{ms}\exp\left(-\frac{IC_2 v^2}{3}t^3\right),$$ (A.5)

where $r_{ms}$ is the meta-stable density of the resident. Recall that we define the meta-stable lifetime $\langle\tau\rangle$ as the time until the residents' density decays to one half of its meta-stable value $r_{ms}$, i.e., $\rho_r(\langle\tau\rangle) = r_{ms}/2$. Then from Eq. (A.5) we obtain

$$\langle\tau\rangle = \left[\frac{3\ln(2)}{IC_2 v^2}\right]^{\frac{1}{3}},$$ (A.6)

which explicitly shows the dependence of the lifetime on the nucleation rate per unit volume $I$ and the radial growth velocity of the invading clusters $v$. Using this expression for the lifetime, we can write Eq. (A.5) in the form of Eq. (4) of the text

$$\rho_r(t) = r_{ms}\exp\left[-\ln(2)\left(\frac{t}{\langle\tau\rangle}\right)^3\right].$$ (A.7)

From the above derivation, Avrami's law, Eq. (A.7), is generic when the switching mechanism is governed by homogeneous nucleation. Parameters of the ecological dynamics ($\alpha_i$, $\beta$, $\mu$ and $\delta$) govern the meta-stable lifetime through their impact on the nucleation rate per unit volume $I$ and the radial growth velocity $v$ [Eq. (A.6)], leaving the functional form of the of the time-dependent density unchanged in Eq. (A.7).

**Figures**

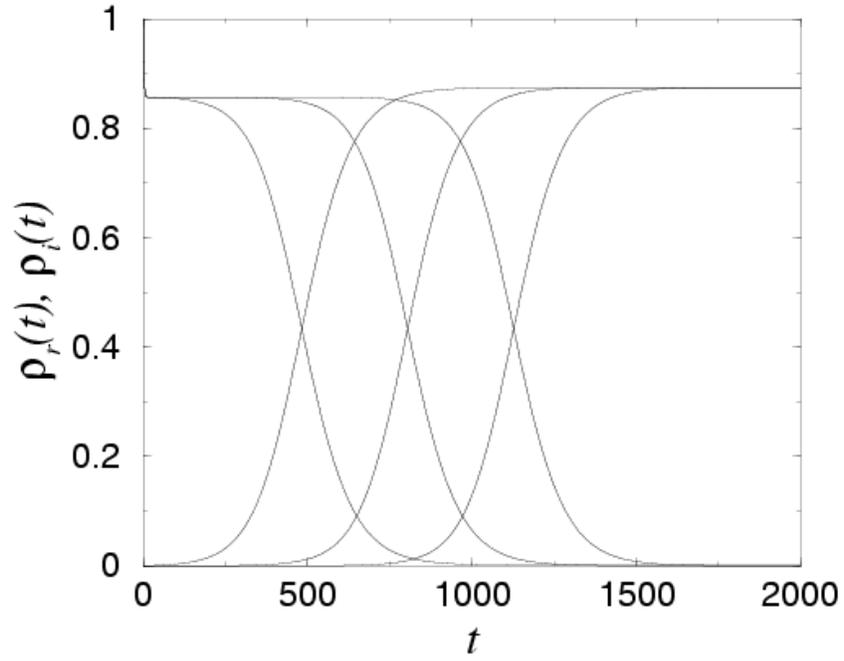

**Fig. 1.** Time-dependent global densities obtained by numerically integrating the mean-field equations for $\beta = 10^{-4}$, $10^{-6}$, and $10^{-8}$ (from left to right, respectively). $\alpha_r = 0.70$, $\alpha_i = 0.80$, and $\mu = 0.10$ throughout this paper. Matching pairs of $\rho_r(t)$ and $\rho_i(t)$ intersect near a density of 0.425.

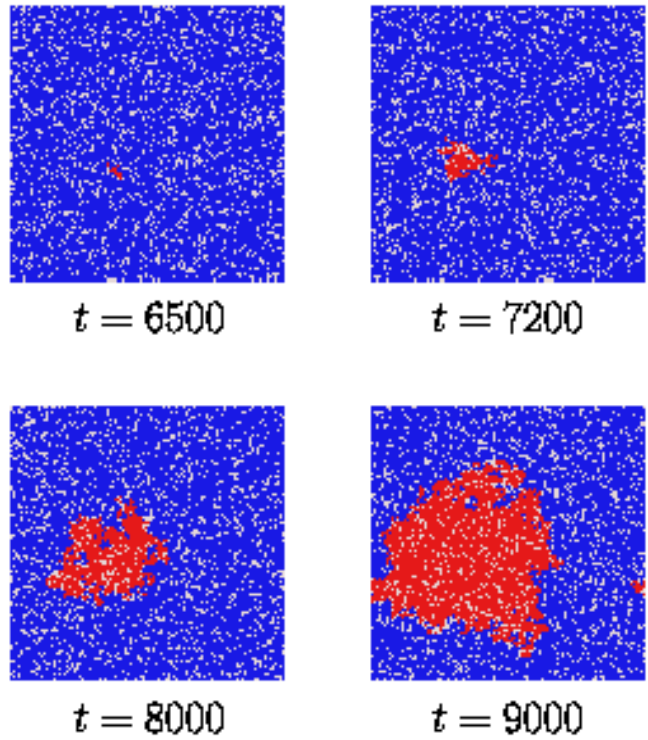

**Fig. 2.** Single-cluster invasion mode in lattice Monte Carlo simulations. Times are given in units of Monte Carlo steps per site (MCSS); $L = 128$ and $\beta = 10^{-6}$. White represents empty sites, blue and red correspond to sites occupied by resident and invasive species, respectively. The waiting time for successful introduction exceeds the time between initiation of invasion and the resident becoming numerically superior.

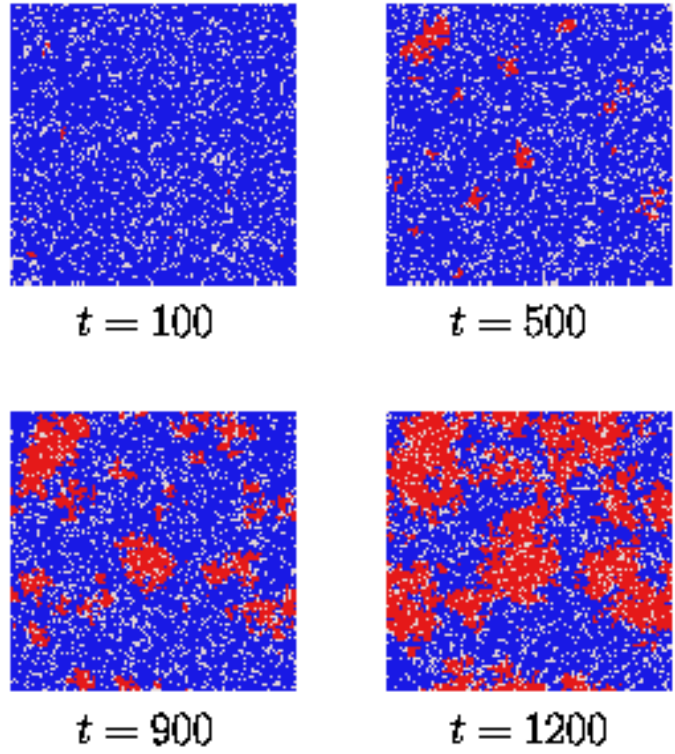

**Fig. 3.** Multi-cluster invasion mode in lattice Monte Carlo simulations. Times are given in units of MCSS; $L = 128$ and $\beta = 10^{-4}$. Sites are color coded as in Fig. 2. The first successful introduction occurs rapidly; additional clusters nucleate during invasion. Note that the system size $L$ is the same as in Fig. 1. The temporal sequence of configurations in Fig. 2 and Fig. 3 demonstrates the importance of the interplay of the characteristic length scales: the typical cluster separation and the system size. In our simulations, critical cluster size is clearly smaller than either underlying length scale.

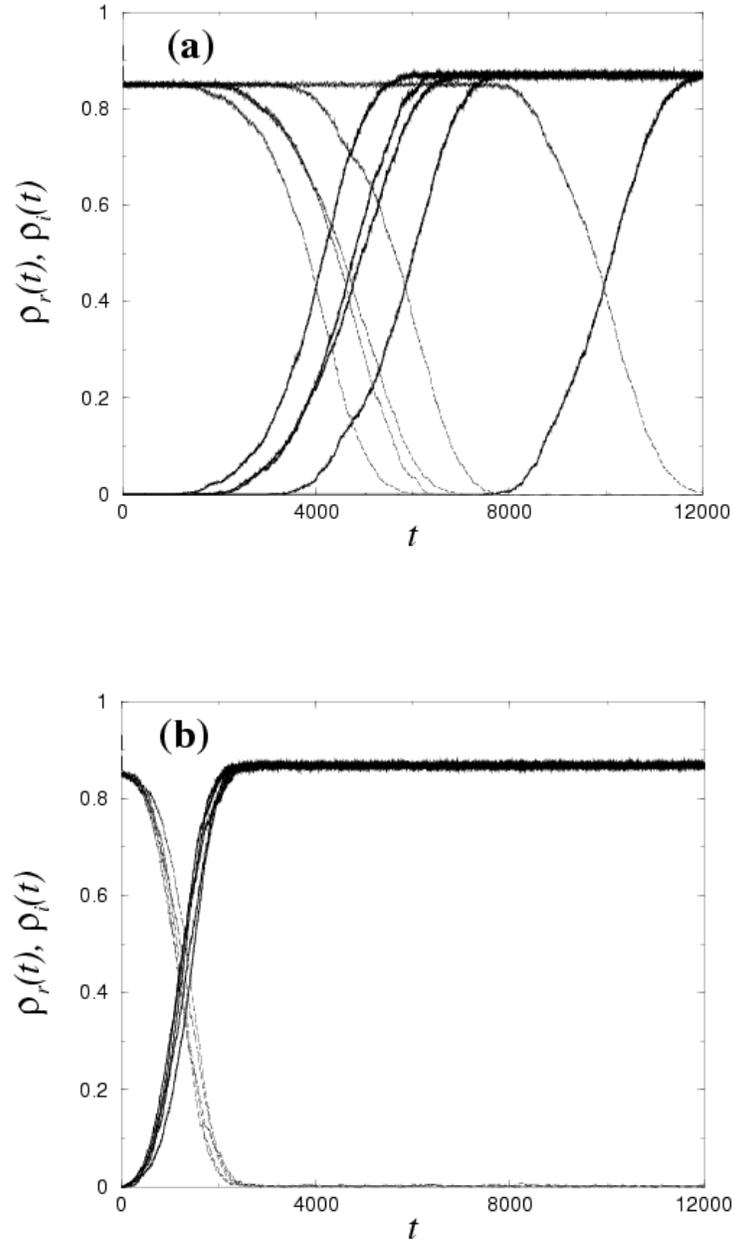

**Fig. 4.** Five independent realizations of the time series of the two species' global densities during (**a**) single-cluster invasion mode. Note that matching pairs of $\rho_r(t)$ and $\rho_l(t)$ intersect near a density of 0.425. (**b**) During multi-cluster (self-averaging) invasion mode. The parameters are the same as those in Fig. 2 and Fig. 3 for (a) and (b) respectively, i.e., $L = 128$ for both, $\beta = 10^{-6}$ in (a) and $\beta = 10^{-4}$ in (b).

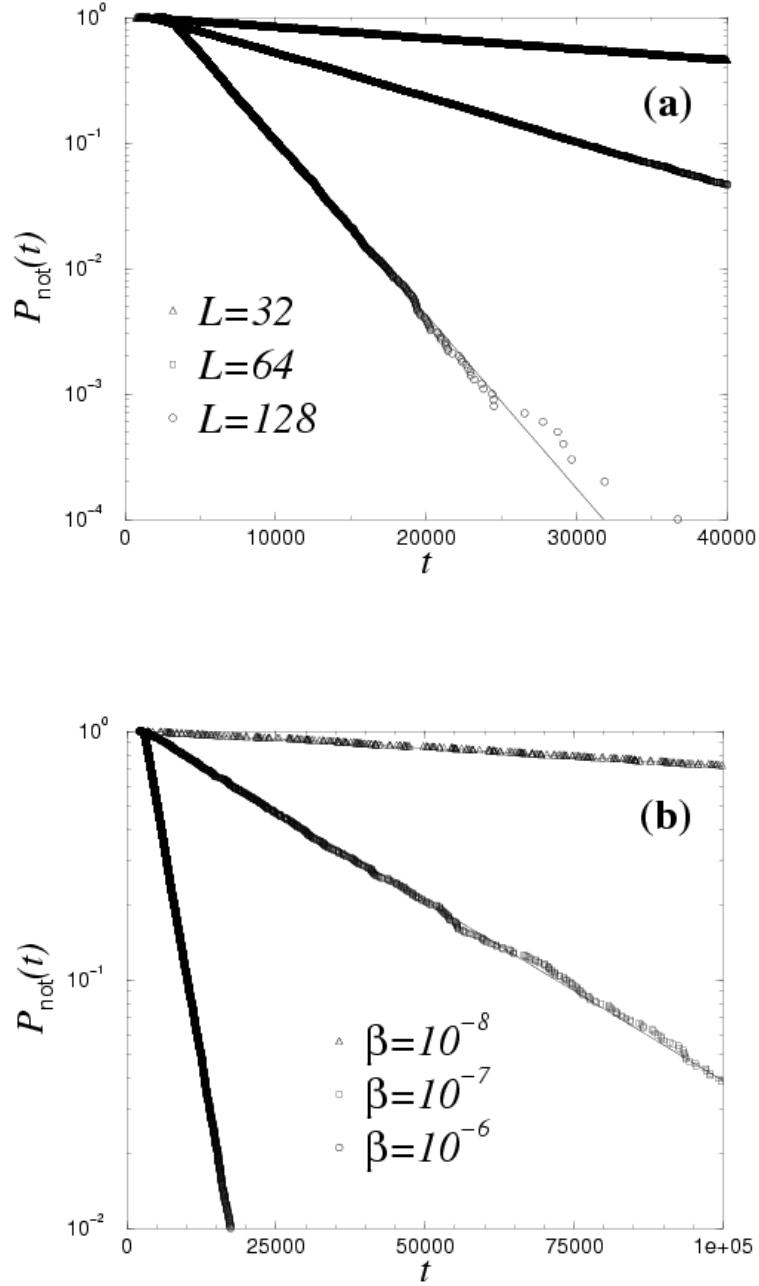

**Fig. 5.** Cumulative probability distribution for the lifetime of the resident species (the probability that the resident's density has not decayed to $r_{ms}/2$ by time $t$) in the single-cluster invasion regime for **(a)** $\beta = 10^{-6}$ and $L$=32, 64, 128 (from top to bottom, respectively); **(b)** for $L$=128 and $\beta = 10^{-8}$, $10^{-7}$, $10^{-6}$ (from top to bottom, respectively). The log-linear scales indicate an exponential distribution for the nucleation process, equation (9). The lifetime distributions were constructed using $10^4$ independent Monte Carlo runs for $\beta = 10^{-6}$; $L$=32, 64, 128 and $10^3$ independent runs for $L$=128; $\beta = 10^{-8}$, $10^{-7}$.

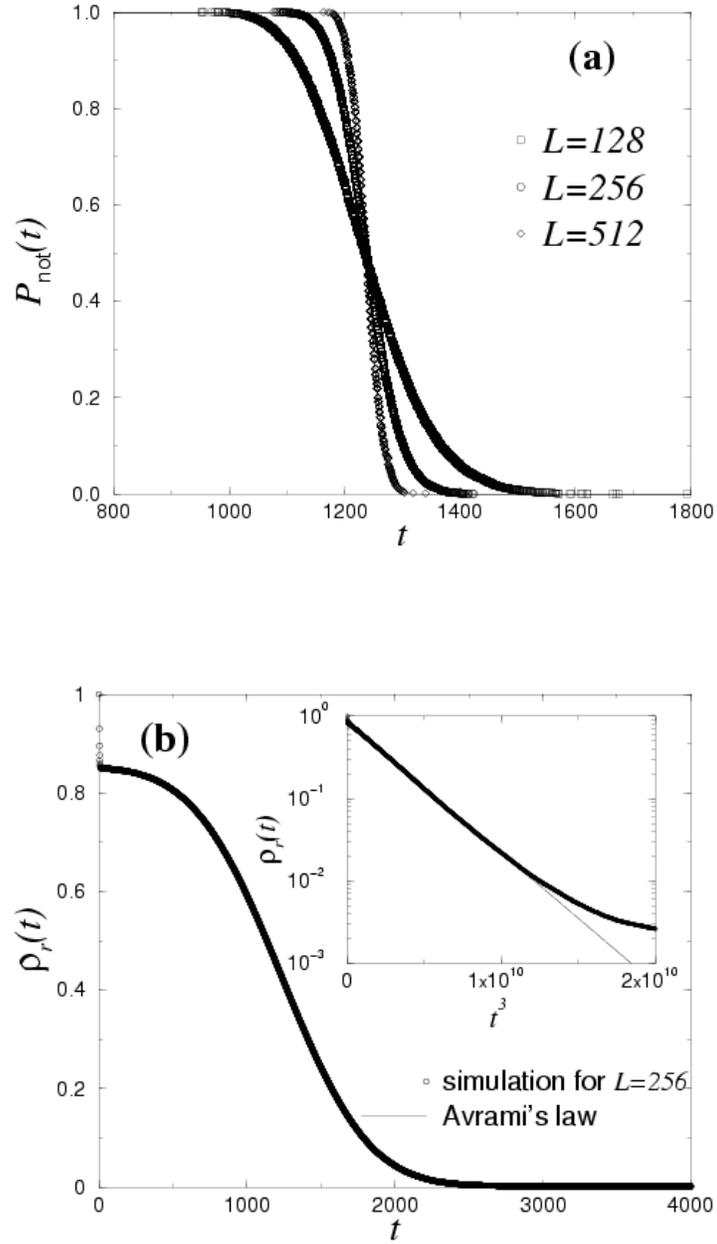

**Fig. 6.** (**a**) Cumulative probability distribution for the lifetime of the resident species (the probability that the resident's density has not decayed to $r_{ms}/2$ by time $t$) in the multi-cluster (self-averaging) invasion with $\beta = 10^{-4}$ for system sizes indicated. Error functions, with the recorded averages and standard deviations as parameters, are plotted over the corresponding simulation data. The lifetime distributions were constructed using $10^4$ independent Monte Carlo runs. (**b**) Time-series of the density of the resident species $\rho_r(t)$ for a $256 \times 256$ system (using an ensemble average of 100 independent runs) for the same value of $\beta$ as in (a). The solid line represents Avrami's law, Eq. (4), not distinguishable from simulation data on these scales. The inset shows $\rho_r(t)$ vs $t^3$ on log-linear scales indicating where the deviation from Avrami's law becomes noticeable.

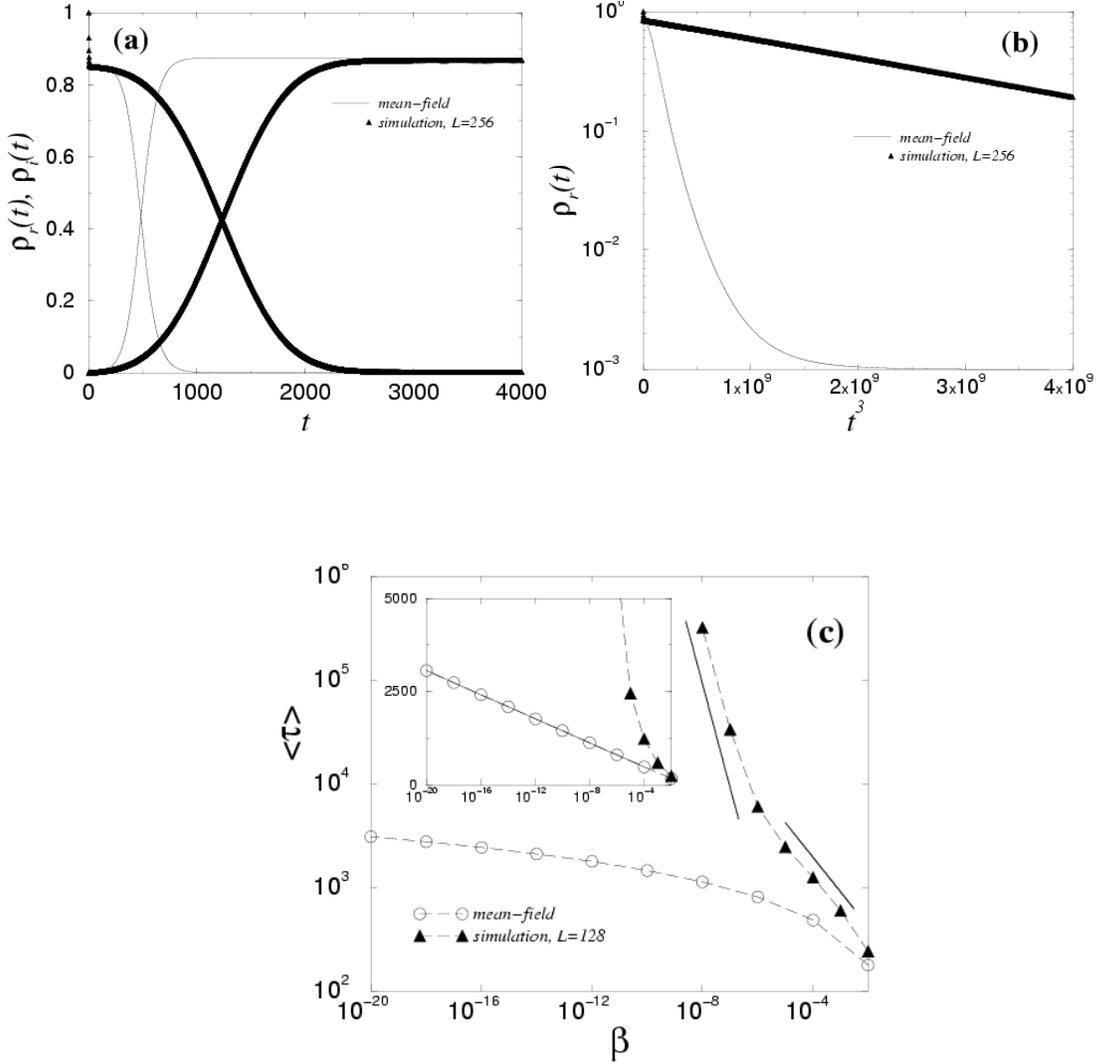

**Fig. 7.** Comparison of the mean-field approximation with lattice Monte Carlo (MC) simulation results. **(a)** Time-dependent global densities for $\beta = 10^{-4}$. Here the system size for MC simulations was $L = 256$, yielding self-averaging multi-cluster invasion. **(b)** The resident's global density vs $t^3$ shown on log-normal scales illustrate the validity of nucleation theory and Avrami's law, and the weakness of the mean-field approximation. **(c)** Lifetime of the resident in the mean-field approximation and by Monte Carlo (MC) simulations for $L=128$ as a function of the introduction rate $\beta$. Log-log scales are used to capture the range of $\beta$ spanning several orders of magnitude and the resulting disparate timescales for the lifetime $\langle\tau\rangle$ in the MC simulations. Standard error for the MC lifetime data is less than 4% for all data points (and would correspond to error bars less than the symbol size for the average lifetime on the graph). The two straight-line segments correspond to the $\beta$-dependence of the average lifetime predicted by nucleation theory in the multi-cluster regime (slope $-1/3$ on log-log plot) and in the single-cluster regime (slope $-1$ on the log-log plot). The inset shows the same data on linear-log scales to illustrate the linear dependence of the lifetime on $\log(\beta)$ for the mean-field approximation. The straight line is the best linear fit, $a - b\log(\beta)$, to the mean-field model.

**Table 1**

List of model symbols, definitions (numerical value or range used in the simulations, where appropriate)

| Symbol | Definition |
| --- | --- |
| $L$ | Lattice length/width ($32 \leq L \leq 512$) |
| $\sigma$ | Set of lattice site's elementary states (empty, invader, resident) |
| $\beta_i$ | Invader's introduction rate ($10^{-8} \leq \beta_i \leq 10^{-2}$) |
| $\beta_r$ | Resident's introduction rate ($\beta_r = \beta_i = \beta$) |
| $\delta$ | Neighborhood size for clonal growth (4) |
| $\alpha_i$ | Invader's clonal propagation rate (0.8) |
| $\alpha_r$ | Resident's clonal propagation rate (0.7) |
| $\eta_i(x, t)$ | Number of invader neighbors around site $x$ at time $t$ |
| $\eta_r(x, t)$ | Number of resident-species neighbors around site $x$ at time $t$ |
| $\mu_i$ | Invader's mortality rate (0.1) |
| $\mu_r$ | Resident's mortality rate ($\mu_i = \mu_r = \mu = 0.1$) |
| $\rho_i$ | Invader's global density |
| $\rho_r$ | Resident's global density |
| $r_{ms}$ | Resident's "metastable" global density |
| $<\tau>$ | Resident's metastable lifetime |
| $t_i$ | Waiting time for invader's nucleation |
| $t_g$ | Time for successful invader to grow to competitive dominance |
| $I(\beta)$ | Nucleation rate per unit area |
| $v$ | Velocity at which cluster radius grows |
| $S(t, t')$ | Volume of cluster at time $t$ formed at time $t' < t$ |
| $r_c$ | Initial radius of nucleating cluster |
| $C_d$ | Dimension-dependent multiplier |
| $d$ | Dimension of volume within which nucleation occurs |

_________________________________________________